# In-Situ Growth of Halide Perovskite Single Crystals and Thin Films on Optical Fiber End Facets

Yang Yu[1], Kanak Kanti Bhowmik[1], Ruan Li[1], Kexin Li[1], Lin Zhu[1], Hai Xiao[1], Lianfeng Zhao[1],*

[1]Holcombe Department of Electrical and Computer Engineering, Clemson University, Clemson, South Carolina 29634, United States

*Email: Lianfez@clemson.edu

## Abstract

Halide perovskites exhibit significant advantages for active optical components such as light emitting diodes, solar cells and photodetectors due to their excellent optoelectronic properties. Their nonlinear optical effects and other characteristics also make them suitable for integration into waveguide components, such as optical fibers, for applications like optical modulation. Although some efforts have been made to integrate perovskite nanomaterials with optical fibers, technological challenges have hindered reliable in-situ preparation methods. Herein, we propose an area-selective wetting strategy for optical fibers, which utilizes hydrophobic sidewalls and hydrophilic end facets to reliably hold small precursor droplets. By introducing a space confinement strategy to suppress the kinetics of solvent evaporation, Methylammonium lead bromide ($MAPbBr_3$) perovskite single crystals were successfully grown in-situ on the fiber end facet. The versatility of this in-situ growth method for single crystals on fiber end facets of various sizes has also been verified. In a separate approach, the controllable in-situ preparation of $CsPbBr_3$ polycrystalline thin films was achieved through vacuum-assisted rapid crystallization. Our strategy provides a controllable platform for the integration of perovskite materials and optical fibers, enabling further development in optical applications.

## 1. Introduction

Halide perovskites are an emerging class of optoelectronic semiconductors. They possess a compelling set of optoelectronic properties, such as a high absorption coefficient,[1] high photoluminescence quantum yield,[2] long carrier diffusion length,[3] and extended radiative recombination lifetime.[4] A key advantage is their compositionally tunable electronic band structure, which allows for precise engineering of their optical and electrical characteristics.[5] These properties make them promising candidates for applications such as solar cells,[6] photodetectors,[7] light-emitting diodes (LEDs),[8] lasers,[9] and neuromorphic devices.[10]

It is noteworthy that beyond photovoltaics, the most significant achievements of halide perovskites in optical applications have been in active optical components like LEDs, and photodetectors. For example, Liao et al.[11] reported a photonic neural network based on a hetero-integrated platform of perovskites and $Si_3N_4$, where the perovskite served as the active layer for both light emission and detection, while $Si_3N_4$ acted as the waveguide. However, passive optical components made from halide perovskites, such as filters, couplers, and splitters, have been less explored. While bottom-up, template-assisted growth has demonstrated polariton propagation in waveguides based on $CsPbBr_3$ nanowires,[12] this approach faces certain limitations. Issues with process compatibility, a lack of universal methods, and constraints on effective refractive indices make heterogeneous integration of perovskites with other optical components a more prevalent strategy.

Optical fiber, as one of today's most important optical components, is widely used in fields such as communication, sensing, and healthcare due to its wide bandwidth, low loss, and resistance to electromagnetic interference.[13] Perovskites, with their strong absorption, efficient electro-optical and photoelectric conversion, and nonlinear effects, are complementary materials for integration into fiber-based systems for detection, emission, modulation, and processing.[14] The solution-processability and compositional diversity of perovskites provide a feasible pathway for realizing integrated perovskite/optical fiber applications. Previous work by Pengfei et al.[15] utilized a dry-transfer method to integrate a methylammonium lead iodide ($MAPbI_3$) single nanosheet onto a fiber end facet, creating a saturable absorber (SA) that achieved stable picosecond-duration soliton mode-locked laser pulses. Similarly, Guobao et al.[16] reported an SA based on $MAPbI_3$ with a strong non-linear response, realizing ultrafast (661 fs) mode-locked laser pulses. Inorganic $CsPbBr_3$ nanocrystals have also been synthesized and drop-cast onto the end face of Er-doped fiber to create a stable dissipative soliton fiber laser.[17] While these methods are effective, they often require pre-synthesis of the material followed by a separate transfer step, which can introduce contamination or mechanical damage. Integration based on side-polished fibers, where a precursor solution is spin-coated directly onto the polished surface, is comparatively straightforward but alters the fiber's cylindrical geometry.[18] The potential of combining perovskite properties with optical fibers for various applications such as medical and industrial sensors,[19] is not yet fully realized due to the limited integration methods currently available.

Most existing work on the optical applications of perovskite-fiber integration utilizes the nonlinear absorption effect of the perovskite, a property that does not require complete controllability or high-quality crystalline coverage. In contrast, using the perovskite as a gain medium for applications like fiber-integrated lasers poses a greater challenge, as it requires the deposition of high-quality crystalline material directly on the fiber end. Herein, we propose a systematic strategy to realize fully controlled in-situ growth of halide perovskites on fiber ends. This idea originated from selectively controlling the wettability of different parts of the optical fiber surface to confine the perovskite precursor solution to the desired end facet.[20] A designed treatment process results in hydrophobic sidewalls and hydrophilic end facets. Building on this droplet fixation, we designed specific crystallization strategies based on the properties of different perovskite compositions. For $MAPbBr_3$, dimethyl sulfoxide (DMSO), a co-solvent with a high boiling point and low evaporation rate, was introduced, and the evaporation kinetics were further controlled using a space confinement method. This approach successfully suppressed the nucleation rate to achieve single-crystal growth at the fiber end. On the other hand, for all-inorganic $CsPbBr_3$, which exhibits low solubility, we achieved high-quality polycrystalline thin films using a vacuum-assisted crystallization strategy. All prepared perovskites exhibited strong and compositionally correct photoluminescence peaks, confirming their quality and indicating their potential for further application as optical gain media integrated on fiber ends.

## 2. Results and Discussion

One challenge in growing halide perovskite on a fiber end is confining the precursor solution to the end facet. We have designed a feasible process to confine solution droplets to the end face and prevent their migration to other areas, such as the fiber sidewall. The entire process is shown in Figure 1a. It aims to make the fiber sidewall hydrophobic while leaving the fiber end hydrophilic. First, the optical fiber is stripped of its outer acrylate coating and cleaved by a diamond blade to obtain an exposed, flat end. Then, the fiber is treated with UV light ozone or oxygen plasma to make the entire silica surface clean and hydrophilic. Subsequently, the goal is to protect the end facet, which has already undergone hydrophilic treatment, so that it is unaffected by the following hydrophobic treatment for the sidewall. Although there are many mature methods for depositing materials on substrates, selectively covering only the fiber end face remains challenging. We developed a method combining dipping and spin-coating operations to selectively apply a protective layer to the fiber end surface. A layer of PMMA solution is spin-coated on a hydrophilic glass surface; the optical fiber is then held vertically and brought into contact with the still-wet liquid film. In this way, only the end surface of the optical fiber is coated with PMMA. For this process, a modified PMMA solution was used; introducing dimethylformamide (DMF), a solvent that evaporates slowly, into toluene and reducing the spin-coating time yields an uncured PMMA film that remains workable for the dipping step. After that, the fiber is placed inside a chamber with trichloro(1H,1H,2H,2H-perfluorooctyl)silane under vacuum. In this vapor-phase deposition environment, the functional silane molecules evaporate and distribute throughout the chamber. The molecule's reactive Si-Cl groups bond to the surface hydroxyl groups on the silica fiber sidewall, forming a durable self-assembled monolayer with the hydrophobic fluorinated alkyl group facing outward. The final procedure is to soak the samples in an organic solvent to dissolve and remove the protective PMMA on the fiber end. With this process, the resultant silica optical fibers can hold polarized solutions on the hydrophilic end surface and prevent these solutions from spontaneously flowing down the hydrophobic sidewalls.

Merely fixing the solution to the end face of the optical fiber is insufficient for growing a single crystal. Such small-sized droplets will evaporate quickly, and the limited solute will rapidly nucleate, forming multiple dispersed small crystals simultaneously. To address this, we referred to the idea of space confinement, a technique commonly used in growing single-crystal thin films.[9,20] This involves covering the solution droplet on the fiber end with a hydrophobic glass slide to reduce the area of the liquid-gas interface and significantly decrease the volatilization rate, thereby achieving reliable growth of perovskite single crystals on the end face, as shown in Figure 1b.

To verify the effectiveness of the area-selective hydrophilic/hydrophobic treatment strategy, we prepared optical fibers with fully hydrophilic, fully hydrophobic, and our target configuration (hydrophilic end face and hydrophobic sidewall). We dropped a 0.9 M $MAPbBr_3$ solution (solvent ratio of DMF: DMSO is 1:4) onto the end facets and allowed them to grow naturally at room temperature under the confinement of a hydrophobic glass slide. Figures 1c, 1d, and 1e show optical microscope images of the three sets of samples from a tilted perspective. A fully hydrophilic optical fiber makes it difficult to control the solution, which tends to be squeezed down the sidewall when the glass is applied. The subsequent evaporation of the solvent leads to the formation of crystals in the area where the solution has spread. For fully hydrophobic optical fibers, crystallization on the sidewall does not occur, but it is also impossible to grow a single crystal at the end. Heterogeneous nucleation of crystals depends on surface energy, and both the hydrophobic end face and the glass cover are thermodynamically unfavorable for the formation of crystal nuclei. Kinetically, the liquid-air interface at the droplet's edge is where the solvent evaporates outward most rapidly, allowing the solution there to reach the supersaturation required for crystallization first. This explains why crystals tend to distribute around the outer circular area of the hydrophobic fiber end. Finally, a single crystal was successfully obtained for the sample with a hydrophilic end face and hydrophobic sidewalls. The hydrophilic end face provides an advantageous nucleation site, the hydrophobic sidewall ensures the solution droplet is confined to the end face, and the hydrophobic glass cover effectively suppresses the rapid evaporation of the solvent. This area-selective wetting strategy successfully demonstrated the ability for controllable growth of a MAPbBr$_3$ single crystal on the fiber end.

MAPbBr$_3$ was selected as the model compound for this study because it is a prototypical halide perovskite with a stable crystal structure and an ideal structural tolerance factor. Its fundamental properties, such as lattice constant and bandgap, are intermediate to those of its chlorine- and iodine-based counterparts.[21,22] Crucially, MAPbBr$_3$ is particularly well-suited for demonstrating single-crystal growth. It exhibits moderate solubility in common organic solvents like DMF and DMSO and possesses an inverse temperature solubility curve that is advantageous for crystallization. These characteristics permit the use of multiple straightforward fabrication techniques, including slow solvent evaporation,[23] inverse temperature crystallization,[24] and anti-solvent induced crystallization.[25] In contrast, MAPbI$_3$ requires high-temperature processing with γ-butyrolactone (GBL),[26] and forms a different, albeit photoactive, tetragonal phase near room temperature, adding experimental complexity.[27] Therefore, the accessible and reliable preparation of MAPbBr$_3$ makes it the ideal candidate for exploring in-situ single-crystal growth on optical fiber end facets.

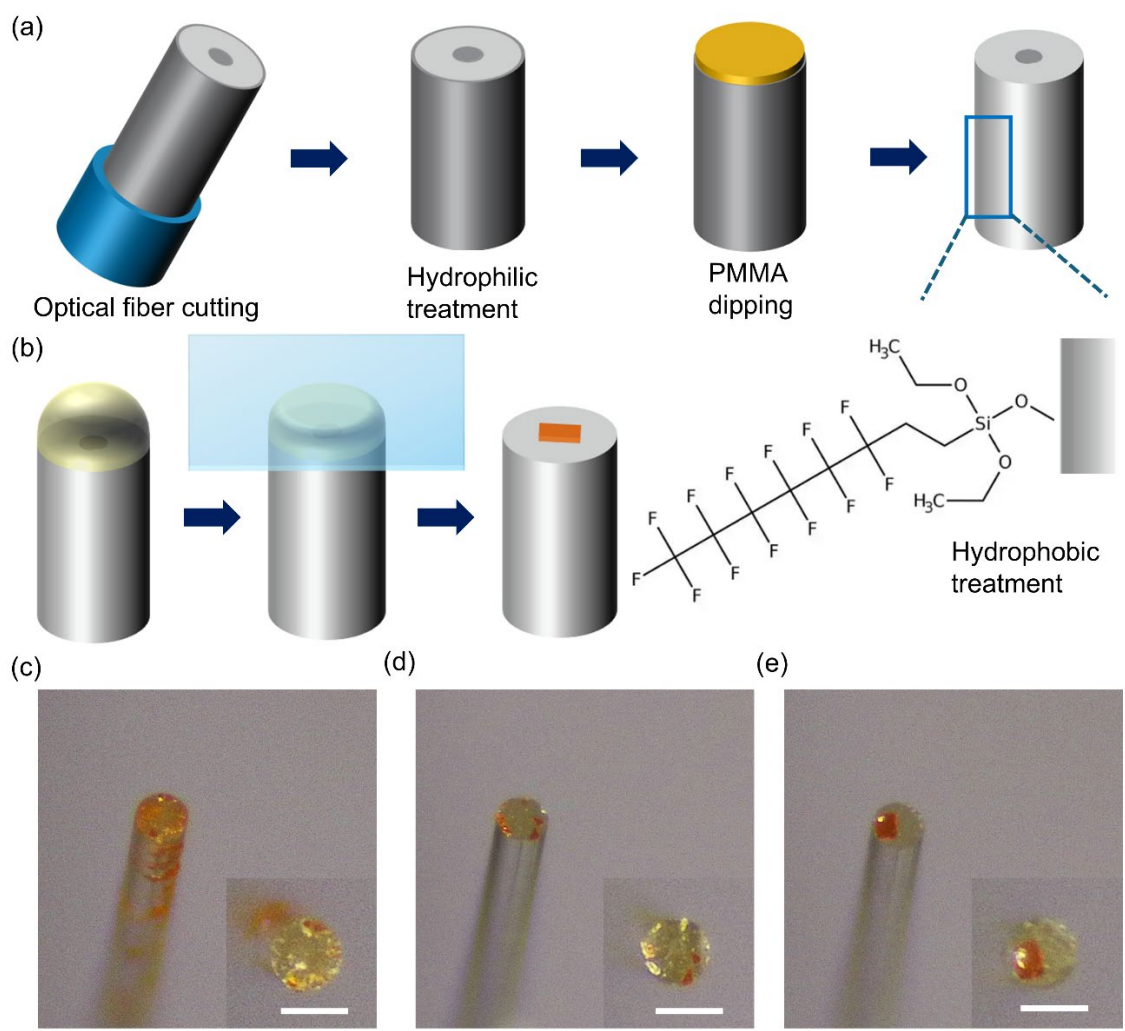

Figure 1. Strategy of hydrophobic sidewall and hydrophilic end facet for growth of MAPbBr$_3$ single crystals. (a) Schematic diagrams of process procedures. (b) Schematic diagrams of space confinement method for single crystal growth on fiber end surface. Optical images of MAPbBr$_3$ crystals on optical fiber with (c) fully hydrophilic surface, (d) fully hydrophobic surface and (e) hydrophilic end and hydrophobic sidewalls. Insets in (c-e) are the top view photos. Scale bar is 125 μm.

To further understand the crystallization kinetics of this area-selective wetting strategy, we further studied the effects of solvent composition and solute concentrations. We started from the principle of crystallization and analyzed the evaporation kinetics under different conditions, solvents, and concentrations, specifically focusing on the relationship between solution concentration and time. The derivation of a simplified model can be found in the Supplementary Information (Note 1). As shown in Figure 2a, there is a significant difference in the concentration change curves between samples with spatial confinement and those without. Without the coverage of the hydrophobic glass, the solution droplets are fully exposed, and the solvent evaporates rapidly from the solution-air interface. Under the effect of space confinement, the concentration of the solution changes very slowly, then increases rapidly near the end of the evaporation process. It can be approximately estimated that the confinement condition delays the volatilization process by several times.

Solvent selection is another decisive factor. Different solvents have different evaporation rates; the evaporation rate for DMF is at least several times higher than that of DMSO. Another effect of solvents on crystallization is the solubility of the solutes. Although the degree of supersaturation directly determines the crystallization process, the maximum soluble concentration can be used to indirectly estimate its timing. The solubility of $MAPbBr_3$ in DMF is less than 2 M, whereas DMSO can dissolve up to 3 M. Therefore, based on volatility and solubility, under the same conditions and initial concentration, the crystallization of $MAPbBr_3$ in DMF will begin significantly earlier than in DMSO.

While our model explored concentration changes by assuming infinite solubility, real-world solutions have limited solubility. Consequently, as the solvent evaporates and the solution concentration reaches a critical point, a phase transition occurs, causing crystals to precipitate. This process of nucleation and growth is well-described by the LaMer model.[28] The crystallization process of a solution normally goes through three stages: stage one, where the concentration gradually increases as the solvent evaporates; stage two, where crystal nuclei form once a critical supersaturation level is reached; and stage three, where existing crystals grow larger as more solute precipitates. As mentioned earlier, space confinement and fully exposed natural volatilization led to completely different evaporation rate curves. As shown in Figure 2b, as the solution evaporates, its concentration increases from an initial $C_0$ to the nucleation point, $C_{min}$. The concentration peaks at $C_{max}$ before decreasing due to solute depletion from crystal growth. Under the condition of space confinement, the significantly reduced evaporation rate greatly delays the beginning of stage two. More importantly, the duration of this nucleation stage also increases, which means that the nucleation rate is greatly decreased, providing the opportunity for the formation of single crystals.

After the end of stage two, the solution will remain at its saturation concentration, and the continuous evaporation of the solvent will induce the existing crystals to grow larger rather than forming new nuclei.

Figure 2c shows the growth of MAPbBr$_3$ single crystals under spatially limited conditions using solutions with different concentrations (from 0.3 M to 1.5 M) and solvent compositions (pure DMF, pure DMSO, and mixtures). The composition of the solvent is the decisive factor in determining whether a single crystal will form, regardless of the solution concentration. In high-boiling-point solvents (pure DMSO or DMSO-rich mixtures), there is a strong tendency to grow a single crystal. In this case, the solution concentration primarily determines the final size of the single crystal, as the amount of solution held on the fiber end face is roughly equal for a given contact angle. For DMF-rich or pure DMF solutions, their rapid evaporation and narrow nucleation window make it almost impossible for the solution to grow single crystals under these conditions. This experimental observation is consistent with the conclusion derived from the evaporation model in the Supplementary Information (Note 1).

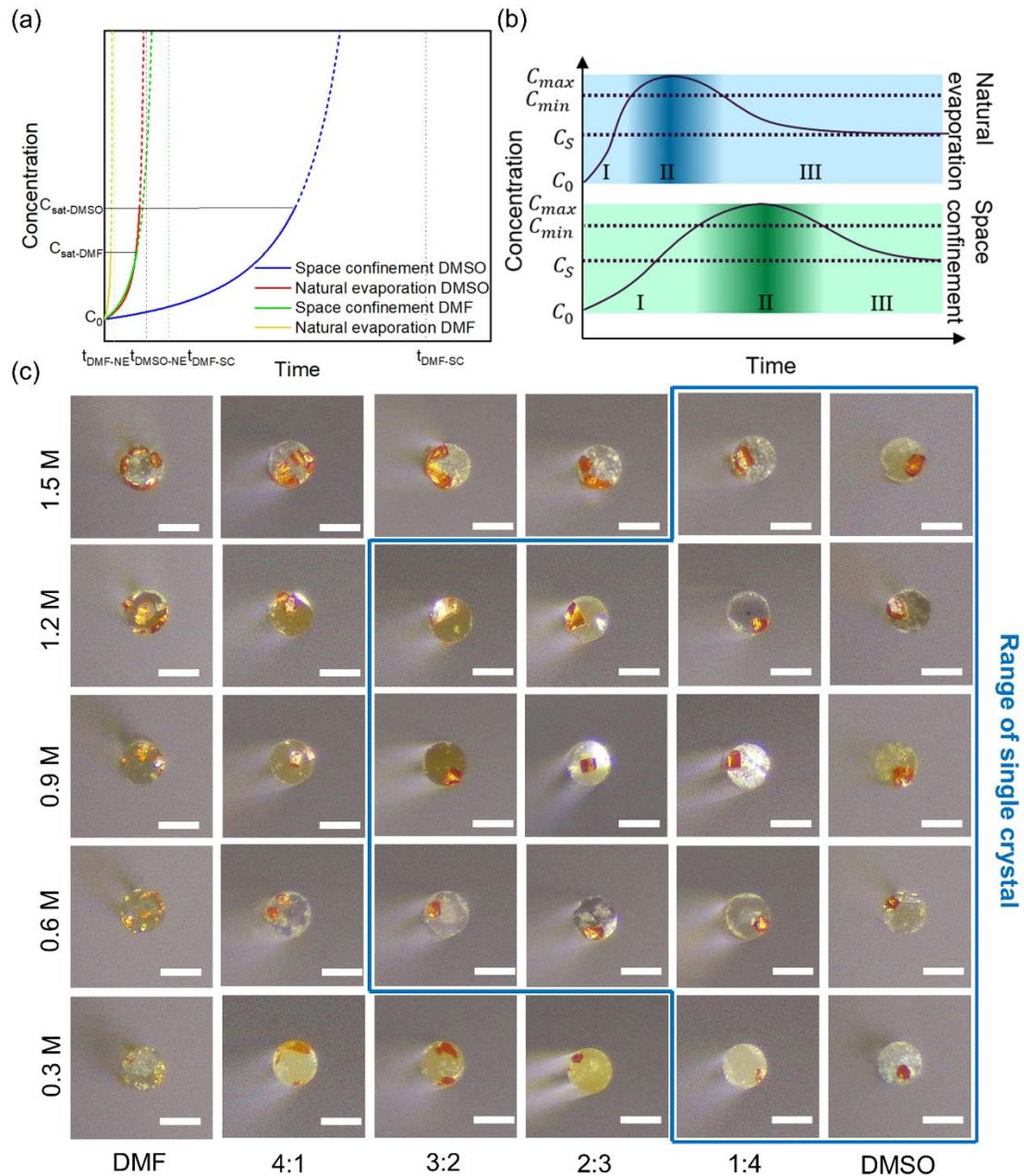

Figure 2. Controlled growth of MAPbBr$_3$ perovskite single crystals on fiber end. (a) Derived relationship between evaporation time and concentration of the solutions under different conditions and various types of solvents. (b) Predicted crystallization stages based on LaMer model for conditions of space confinement and natural evaporation. (c) Optical images of crystallization results for different ratios of DMF and DMSO solvents and concentrations ranging from 0.3 M to 1.5 M. Scale bar is 125 μm.

The material composition of the grown MAPbBr$_3$ was further characterized by scanning electron microscopy (SEM) and photoluminescence spectroscopy. The SEM image (Figure 3a) shows the complete crystal morphology of the MAPbBr$_3$ single crystal. Although the circular fiber end face limits its growth dynamics, preventing the formation of a perfect cube, its surface is smooth and continuous, showing no evidence of grain boundaries. The energy-dispersive X-ray spectroscopy (EDS) mapping in Figure 3b demonstrates the correct, strong, and uniform distribution of the characteristic elements Pb and Br, while the exposed surface of the optical fiber only contains signals of Si and O. Finally, its photoluminescence curve, shown in Figure 3c, exhibited a peak at around 535 nm, consistent with previous literature reports,[9] thereby verifying the composition of the MAPbBr$_3$ single crystal prepared at the fiber end facet.

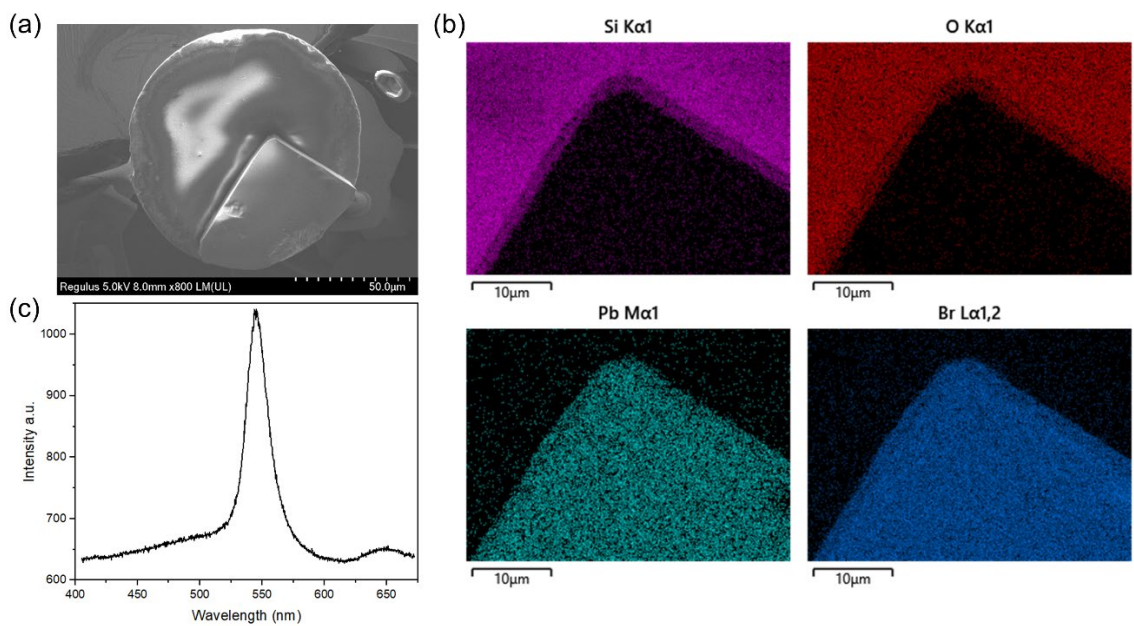

Figure 3. Material characterization of MAPbBr$_3$ single crystal on fiber end. (a) SEM image, (b) EDS mappings and (c) PL spectrum of the MAPbBr$_3$ single crystal.

The optical fibers we studied so far have a diameter of 125 μm. To verify the reliability of our method for preparing perovskite on fiber end facets with different diameters, we used 105, 200, 400, and 600 μm multimode visible light fibers for validation. As shown in Figure 4, MAPbBr$_3$ single crystals can be grown on all the above-mentioned sizes of optical fibers, which paves the way for the subsequent integration of perovskite-based optics with a wider range of optical fibers.

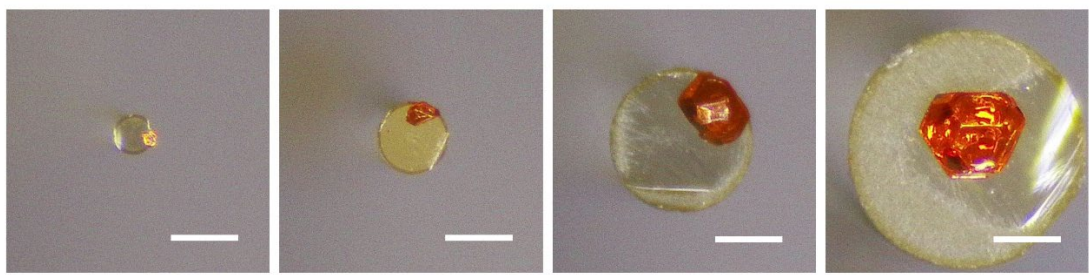

Figure 4. Optical photos of MAPbBr$_3$ single crystals grown on visible light optical fibers with diameters of 105, 200, 400 and 600 μm (from left to right. scale bar: 200 μm).

While controlled solvent evaporation is effective for growing perovskite single crystals, this method makes it challenging to achieve full, continuous coverage on the fiber end facet. For applications that require a complete layer, we therefore shifted our focus to fabricating high-quality polycrystalline thin films using a different approach. The formation of such films depends on inducing high-density nucleation across the substrate in a very short time. To explore this, we used the all-inorganic perovskite CsPbBr$_3$. The principles guiding both the natural and vacuum-assisted growth methods are illustrated in Figure 5a. All-inorganic perovskite precursors generally exhibit low solubility in polar organic solvents. Consequently, a saturated solution of these precursors is highly sensitive to changes in physical conditions, readily becoming supersaturated and promoting a high nucleation rate.[29] When left to evaporate naturally, the solution generated many small, orange crystals of varying sizes instead of a uniform film, as shown in Figure 5b. To achieve a continuous film, the solution must be brought to a state of high supersaturation quickly. Although this can be done by changing temperature or pressure, uniformly heating the small fiber end is challenging. Therefore, we adopted a vacuum-assisted strategy to rapidly reduce pressure, providing an effective method for inducing the fast, widespread crystallization needed for thin film growth. As shown in Figure 5c, under vacuum-assisted conditions, the solution rapidly crystallizes to form a uniform and continuous yellow film, which covers most of the fiber end facet. Figure 5d shows the SEM image of the $CsPbBr_3$ thin film prepared on the surface of the fiber end facet, revealing distinct, large, compact, and continuous perovskite grains. The EDS mapping of the corresponding area in Figure 5e shows clear, strong signals of the characteristic elements Pb and Br. Finally, the photoluminescence spectrum of the sample was tested, showing a strong peak at around 545 nm (Figure 5f), consistent with literature reports and demonstrating the good crystallinity of the thin film.[20]

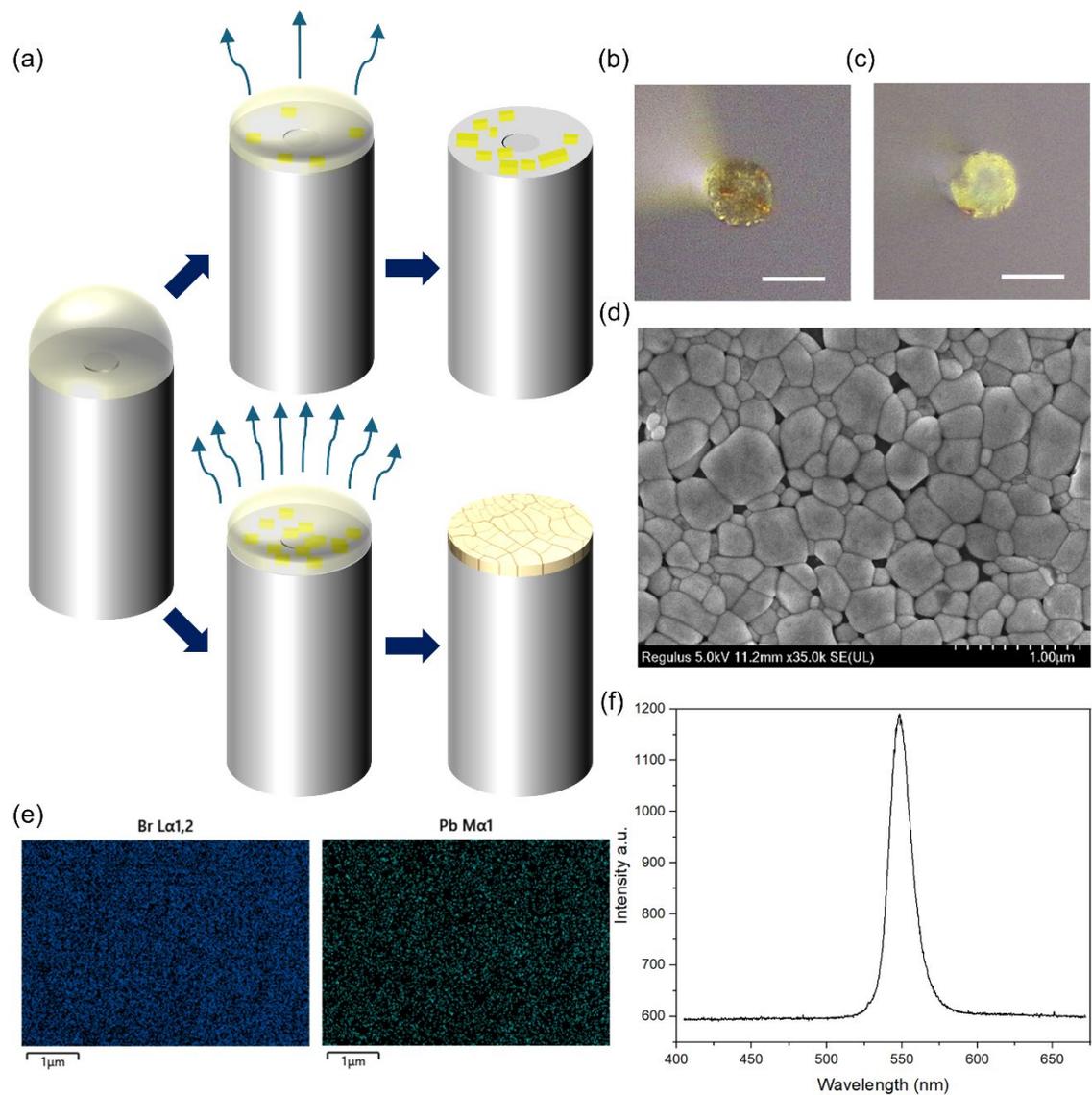

Figure 5. Controllable growth of CsPbBr$_3$ thin film on optical fiber end under vacuum condition and its material characterization. (a) Schematic diagrams of the principle of the vacuum assisted growth method. Optical images of (b) sample grown by naturally evaporated solution and (c) vacuum assisted growth of CsPbBr$_3$ thin film on fiber ends. (d) SEM image of the as-prepared CsPbBr$_3$ thin film on fiber. (e) EDS mapping of elements Br and Pb of the corresponding area of CsPbBr$_3$ thin film in (d). (f) PL spectrum of the CsPbBr$_3$ thin film on fiber end.

## 3. Conclusion

We have demonstrated a method to confine precursor solutions on fiber end facets by implementing an area-selective wettability strategy, which creates hydrophobic sidewalls and a hydrophilic end. This approach enabled the successful in-situ growth of MAPbBr$_3$ perovskite single crystals when combined with a space confinement technique using a hydrophobic glass cover. A comparison of crystal growth under natural evaporation and space confinement was conducted by combining a simplified evaporation kinetics model with the LaMer model of crystallization. This analysis clarified the distinct growth behaviors observed in different solvent systems. The analysis revealed that slow evaporation, facilitated by high-boiling-point solvents under space confinement, is the decisive factor in successfully growing perovskite single crystals from the small-volume droplets held on the fiber end facet. In a complementary approach, the reliable in-situ growth of all-inorganic CsPbBr$_3$ polycrystalline thin films was achieved by employing a vacuum-assisted rapid crystallization method. The high quality of these films was confirmed by corresponding material characterizations, showing dense, continuous grain structures and strong photoluminescence. In summary, this work reports a fully controllable and versatile method for the in-situ growth of either perovskite single crystals or polycrystalline thin films on the end facets of optical fibers. This advancement in direct functional material integration with optical fibers provides a foundational step for developing novel fiber-based optoelectronic devices and systems.

**Experimental Section**

**Materials:** Methylammonium bromide (MABr) was purchased from Greatcell Solar Materials. Lead bromide (PbBr$_2$) was purchased from Tokyo Chemical Industry. Cesium bromide (CsBr), poly(methyl 2-methylpropenoate) (PMMA) and trichloro(1H,1H,2H,2H-perfluorooctyl)silane (PFOCTS), dimethylformamide (DMF), dimethyl sulfoxide (DMSO) and toluene were purchased from Sigma-Aldrich. All materials were used as received without further purification.

**Solutions Preparation:** The MAPbBr$_3$ precursor solution was prepared by dissolving equimolar amounts of MABr and PbBr$_2$ powders in a solvent system consisting of DMF, DMSO, or their mixtures inside a glass vial. The volume ratios of DMF to DMSO investigated were 0:5, 1:4, 2:3, 3:2, 4:1, and 5:0. A range of solution concentrations from 0.3 M to 1.5 M was prepared for the study. The CsPbBr$_3$ precursor solution was prepared by dissolving CsBr and PbBr$_2$ in a 1:1 molar ratio in 1 mL of DMSO to achieve a final concentration of 0.2 M. The PMMA solution was prepared by dissolving 30 mg of PMMA in a 1:1 volume ratio of toluene and DMF. All solutions were ultrasonically vibrated to ensure complete dissolution and were subsequently filtered through a 0.2 μm PTFE membrane filter. All solution preparation and handling steps were performed inside a nitrogen-filled glove box.

**Area-Selective Wettability Treatment of Optical Fibers:** Optical fibers were stripped of their acrylic outer layer using fiber stripping pliers and then cleaved with a diamond fiber cleaver to obtain a flat end facet. The fibers were subsequently cleaned by sequential immersion and sonication in acetone and isopropanol for 10 minutes each. To render the entire silica surface hydrophilic, the fibers were treated with UV-ozone or oxygen plasma. Next, the PMMA solution was spin-coated onto a hydrophilic glass slide at 1000 RPM for 5 s. While the resulting film was still wet, an optical fiber was held vertically with its end facet oriented downwards and was brought into contact with the PMMA solution on the glass. Afterward, the fiber optic sample was placed in a glass desiccator connected to a vacuum pump. Ten microliters of PFOCTS were introduced into the chamber, which was then evacuated for 1 hour to perform a vapor-phase hydrophobic treatment. Finally, the optical fibers, now possessing hydrophobic side walls, were soaked in toluene for several minutes to completely remove the PMMA protective layer from their end surfaces.

**In-Situ Growth of Perovskites on Fiber End Facet:** All operations were monitored using a stereomicroscope equipped with a homemade optical fiber mounting stage. For $MAPbBr_3$ single crystal growth, a glass slide that had undergone hydrophobic treatment with PFOCTS was used for confinement. The fiber was fixed in the mount with its end facet facing vertically upwards. A small droplet of the precursor solution was transferred to the fiber end facet using a pipette tip. Immediately thereafter, a displacement platform holding the downward-facing hydrophobic glass slide was positioned over the fiber and slowly lowered until the glass made contact with the droplet. The sample was then left undisturbed, and crystallization occurred naturally through solvent evaporation in ambient conditions. For $CsPbBr_3$ polycrystalline thin films, the fiber was placed vertically upwards inside a vacuum drying oven. The $CsPbBr_3$ solution was dropped onto the fiber end, and the chamber was evacuated to 0.02 MPa for a minimum of 20 minutes to induce rapid crystallization.

**Material Characterizations:** Scanning electron microscopy (SEM) imaging and energy-dispersive X-ray spectroscopy (EDS) analysis were performed using a Hitachi Regulus 8230 Ultra-High-Resolution SEM. Photoluminescence (PL) spectra were acquired using a custom-built optical platform. The system included a PHAROS PH2 femtosecond laser operating at an excitation wavelength of 257 nm and a Horiba iHR320 spectrometer for photoluminescence detection.

## Acknowledgement

The authors thank Prof. Liang Dong at Clemson University for his insightful discussions. This work is partially supported by the Clemson University Faculty Startup Fund and the National Science Foundation under Award Nos. DMR-2403802 and ECCS-2304364.

## Conflict of Interest

The authors declare no conflict of interest.

# Supplementary Information

## In-Situ Growth of Halide Perovskite Single Crystals and Thin Films on Optical Fiber End Facets

Yang Yu[1], Kanak Kanti Bhowmik[1], Ruan Li[1], Kexin Li[1], Lin Zhu[1], Hai Xiao[1], Lianfeng Zhao[1,*]

[1]Holcombe Department of Electrical and Computer Engineering, Clemson University, Clemson, South Carolina 29634, United States

*Email: Lianfez@clemson.edu

### Note 1: Geometric Model of Solvent Evaporation Kinetics

According to classical nucleation theory[1], the nucleation rate ($J$), number of nuclei per unit time per volume is defined as

$$J = K exp(-\frac{\Delta G^*}{kT}) \quad (1)$$

Where $K$ is the prefactor related to molecular dynamics, $k$ is the Boltzmann constant and $T$ is the absolute temperature. For a certain solution system under a given temperature, the $J$ is determined by the critical nucleation barrier ($\Delta G^*$), which is the change of Gibbs free energy of the system when the crystal nucleus reaches the critical nucleation radius. The $\Delta G^*$ is derived as

$$\Delta G^* = \frac{16\pi\gamma^3 v^2}{3k^2 T^2 (lnS)^2} \quad (2)$$

Where $\gamma$ is the interfacial energy between the solution and solute molecules, $v$ is the volume of the molecule, and $S$ is the supersaturation of the solution. Therefore, in a specific solution system and temperature, the nucleation rate of the solution is only determined by the degree of supersaturation. For a solution that precipitates crystals by solvent evaporation with concentration increasing, the rate of solvent evaporation determines the nucleation rate. To understand why the space confinement strategy on the fiber end facet can reliably achieve single crystal growth, it is necessary to introduce geometric models to compare it with naturally evaporating droplets.

The droplet with a radius of r in the shape of a spherical cap completely covers the end facet of the optical fiber and spreads out to form an approximately circular disk shape when covered with hydrophobic glass. The thickness of the disk is $h$. Therefore, its volume is $\pi r^2 h$ and the area of sidewall is $2\pi r h$. The solvent only evaporates outward on the sidewall in contact with air, and the volume of the solution decreases with time, geometrically manifested as a decrease in the contraction radius of the disk. Thus, its volume change per unit time is

$$dV = 2\pi rh dr \tag{3}$$

And the volume change per unit time is

$$\frac{dV}{dt} = -2\pi rhE \tag{4}$$

Where $E$ is the evaporation coefficient of the solvent. Plug $dV$ in it we obtain

$$dr = -Edt \tag{5}$$

Integrating it can yield

$$r(t) = r_0 - Et \tag{6}$$

Where $r_0$ is the initial radius of the droplet, in fact, it is the radius of the optical fiber. Furthermore, we can plug it in to obtain the concentration formula

$$c = \frac{n}{V} = \frac{n}{\pi r(t)^2 h} = \frac{n}{\pi(r_0 - Et)^2 h} \tag{7}$$

Where $n$ is the amount of substance in the droplet. The volume of liquid droplets that can be bound by the fiber end facet is constant for solutions of the same concentration and solvent, determined by the wettability (contact angle) of the end face surface. In terms of time, before the crystal grows out, the solution can be considered essentially homogeneous, so it makes sense to assume that $n$ remains constant.

For naturally evaporated droplet without confinement, its shape is a spherical cap with a body radius of $R$, a base radius of $r$ and height of $h'$, which are determined by the contact angle of the fiber end facet

$$R = \frac{r}{sin\theta} \tag{8}$$

$$h' = R(1 - cos\theta) \tag{9}$$

Due to the absence of hydrophobic glass, the solution would adhere to the hydrophilic end surface. As the solvent evaporates through the interface of air on the surface of the ball cap, the height of the ball cap will decrease with a constant radius $r_0$. Based on this assumption, the geometric variable under this condition is the angle of liquid ball cap and the end face chamfer, which would gradually decrease as solvent evaporating. However, the volume of the ball cap is $\pi h'^2(3R - h')/3$, and it will be very difficult to calculate the numerical value when $R$ and $h'$ is substituted by $\theta$. Our goal is to make a qualitative comparison, and herein we have made an approximation based on the actual situation. Considering the initial angle, the contact angle of the end facet should be less than 10°, and it would continue to decrease as the liquid evaporates. Therefore, these geometric variables at small angles can be approximated as

$$R \approx \frac{r_0}{\theta} \tag{10}$$

$$h' \approx r_0 tan\theta \approx r_0\theta \tag{11}$$

Therefore, the exposed area of the droplet is

$$A' = 2\pi R h' \approx 2\pi r_0^2 \tag{12}$$

And the volume of the droplet (h << R) is

$$V' \approx \pi h'^2 R \approx \pi \theta r_0^3 \tag{13}$$

By substituting them into the evaporation formula, we can obtain

$$\frac{dV'}{dt} = \frac{3\pi r_0^3 d\theta}{dt} = -2\pi r_0^2 E \tag{14}$$

Cancelling $\pi r_0^2$ yields

$$d\theta = -\frac{2E}{3r_0} dt \tag{15}$$

Integrating the formula yields

$$\theta = \theta_0 - \frac{2E}{3r_0}t \tag{16}$$

Where $\theta_0$ is the contact angle of the fiber end surface. Thus, the concentration is

$$c' = \frac{n}{V'} = \frac{n}{\pi\theta r_0^3} = \frac{n}{\pi(\theta_0 - \frac{2E}{3r_0}t)r_0^3} \tag{17}$$

Since we have equations (7) and (17), by making the denominators equal to zero, we can obtain the termination time when the solution completely evaporates. For space confinement growth

$$t = \frac{r_0}{E} \tag{18}$$

For naturally evaporated droplet

$$t' = \frac{3\theta_0 r_0}{2E} = \frac{3\theta_0}{2}t \tag{19}$$

Because $\theta_0 < 10° \approx 0.1745$, $t$ should larger than about $4t'$. It means that for a certain volume of solution, the evaporation time of the solvent in the spatial confinement method is at least four times that of natural exposure evaporation. For a more realistic scenario, the contact angle of completely hydrophilic silica is approximately 5 °, so this difference is even greater. The formula (17) has a steeper shape with a greater slope in the area near the asymptote. Therefore, for a determined initial concentration, the rate of concentration increase in the solution under spatial confinement is very slow, resulting in a low nucleation rate. If the evaporation coefficient of the solvent is introduced, the curve of DMSO, a high boiling point solvent, will be very smooth in the early stage, that support the growth of single crystal eventually. These simplified geometric models reliably validate the theoretical trend of controllable growth.